\documentstyle[twoside,fleqn,espcrc2,psfig]{article}

\newcommand{\AmS}{{\protect\the\textfont2
  A\kern-.1667em\lower.5ex\hbox{M}\kern-.125emS}}
\title{Fast Fermion Monte Carlo
\thanks{Talk presented by T. Takaishi at LATTICE96}}

\author{Philippe de Forcrand\address{Swiss Center for Scientific Computing, 
        ETH-Zentrum, RZ,   
        8092 Z\"urich, Switzerland}
        and 
        Tetsuya Takaishi$^{\rm a}$
}

\begin{document}

\begin{abstract}
Three possibilities to speed up the Hybrid Monte Carlo algorithm are 
investigated. Changing the step-size adaptively brings no practical gain.
On the other hand, substantial improvements result from using an approximate 
Hamiltonian or a preconditioned action.
\end{abstract}

\maketitle

\section{Introduction}
The Hybrid Monte Carlo (HMC) algorithm\cite{HMC} is now 
the standard method to simulate dynamical fermions. 
However the cost of the algorithm is substantial.
It is therefore desirable to improve the efficiency 
of the HMC algorithm.

First, we examine the possibility of changing the step size
adaptively\cite{Forcrand,adapt}. 
Second, we implement the HMC algorithm with 
approximate Hamiltonians. 
Lastly, we precondition the fermionic action used in the HMC algorithm.

\section{HMC with adaptive step size}
The conventional HMC algorithm is performed with a fixed step size.
One expects however that the molecular dynamics (MD) trajectory will bounce 
off the effective energy barrier represented by the minimum of the
fermionic determinant. As it does, the curvature increases and so
does the integration error. This effect should be more pronounced
for smaller quark masses. Therefore one might expect, in that regime,
benefits from varying the step size adaptively.

The naive way to adapt the step size consists in keeping the local error
constant at each integration step. But care must be taken to preserve
time-reversibility.
Stoffer has constructed a time-reversible adaptive step size method
which gave promising gains on the Kepler problem\cite{Stoffer}. 
Here we implement this method for the HMC algorithm.

Call $T(\Delta t)$  a one-step integrator;
$T(\Delta t):(p,U) \longrightarrow (p^\prime,U^\prime)$.
It maps momenta and link variables $(p,U)$ onto $(p^\prime,U^\prime)$.
Now we define a symmetric error estimator,
\begin{equation}
E_S (p,U:\Delta t) = e(p,U:\Delta t) + e(p^\prime, U^\prime : -\Delta t),
\label{ES}
\end{equation}
where $e(p,U:\Delta t)$ is a local error at $(p,U)$ when the system is integrated by 
$T(\Delta t)$.
If the integrator is reversible, eq.(\ref{ES}) is obviously symmetric under the exchange
$(p,U,\Delta t) \longleftrightarrow (p^\prime, U^\prime,-\Delta t)$.
The adaptive step size $\Delta t$ is then determined by solving a symmetric error equation,
\begin{equation}
\label{tol}
E_S(p,U:\Delta t) = tolerance
\end{equation}
The tolerance should be kept constant during the MD simulation.
The adaptive step size determined by eq.(\ref{tol}) takes
the same value at the reflected point $(-p^\prime, U^\prime)$.
Therefore an integrator with the adaptive step size determined by eq.(\ref{tol})
is reversible.
Taking for $T$ the standard leapfrog integrator,
we tested the above method for several quark masses, volumes and couplings\cite{adapt}. 
We observed that the variance of the step size increases as $\kappa$ increases,
as expected.
However, it decreases as the volume increases.
Thus the gain over HMC with fixed step size turns out to be very small. 
The overhead to determine the adaptive step size by solving eq.(\ref{tol})
exceeds this gain.

\section{HMC with approximate Hamiltonian}
The leapfrog integrator with step size $\Delta t$ has 
$O(\Delta t^3)$ errors, resulting in an energy 
violation $\Delta H$, which reduces the Metropolis acceptance.
Since such errors are present anyway, there is no need to keep 
for the molecular dynamics Hamiltonian $H_{MD}$
the exact Hamiltonian $H$.
An approximate Hamiltonian $H_{MD}$ which introduces errors of similar magnitude
can be substituted at cheaper cost.
The cost of the HMC algorithm is expressed by the cumulated cost of fermionic
force evaluations per accepted trajectory:
\begin{equation}
C_t =  \frac{N_D}{Acc\times \Delta t}
\end{equation}
for trajectories of a certain fixed length, where $N_D$ is the number of multiplications
by the Dirac operator $D$ at each step of size $\Delta t$.
In the following, we consider two possibilities to decrease $N_D$
and try to find the minimum cost $C_t$.
 
\subsection{Chebyshev polynomial approximation to the Hamiltonian}
\begin{figure}[htb]
\vspace{9pt}
\centerline{\hbox{
\psfig{figure=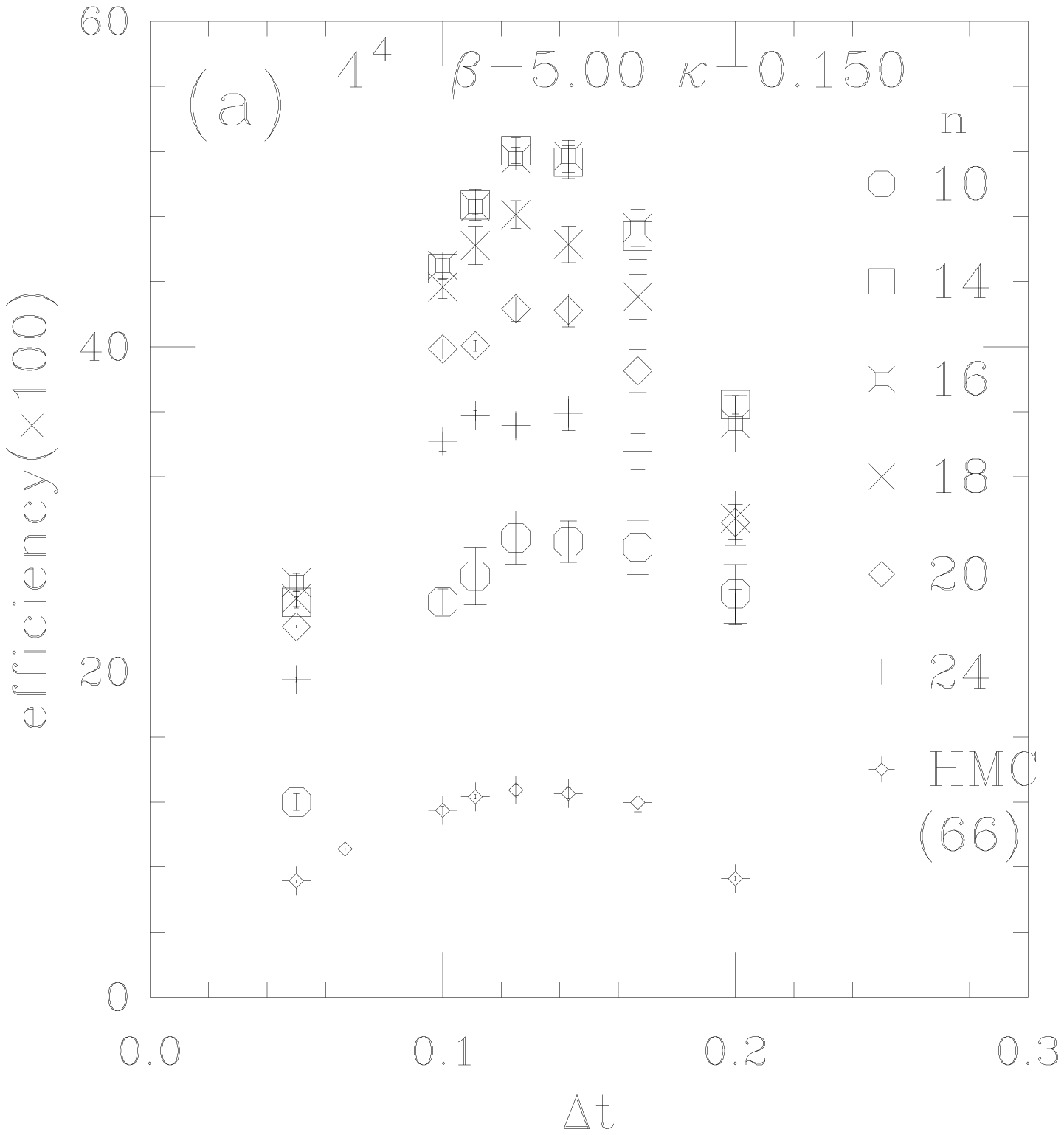,height=3cm,width=4.cm} 
\psfig{figure=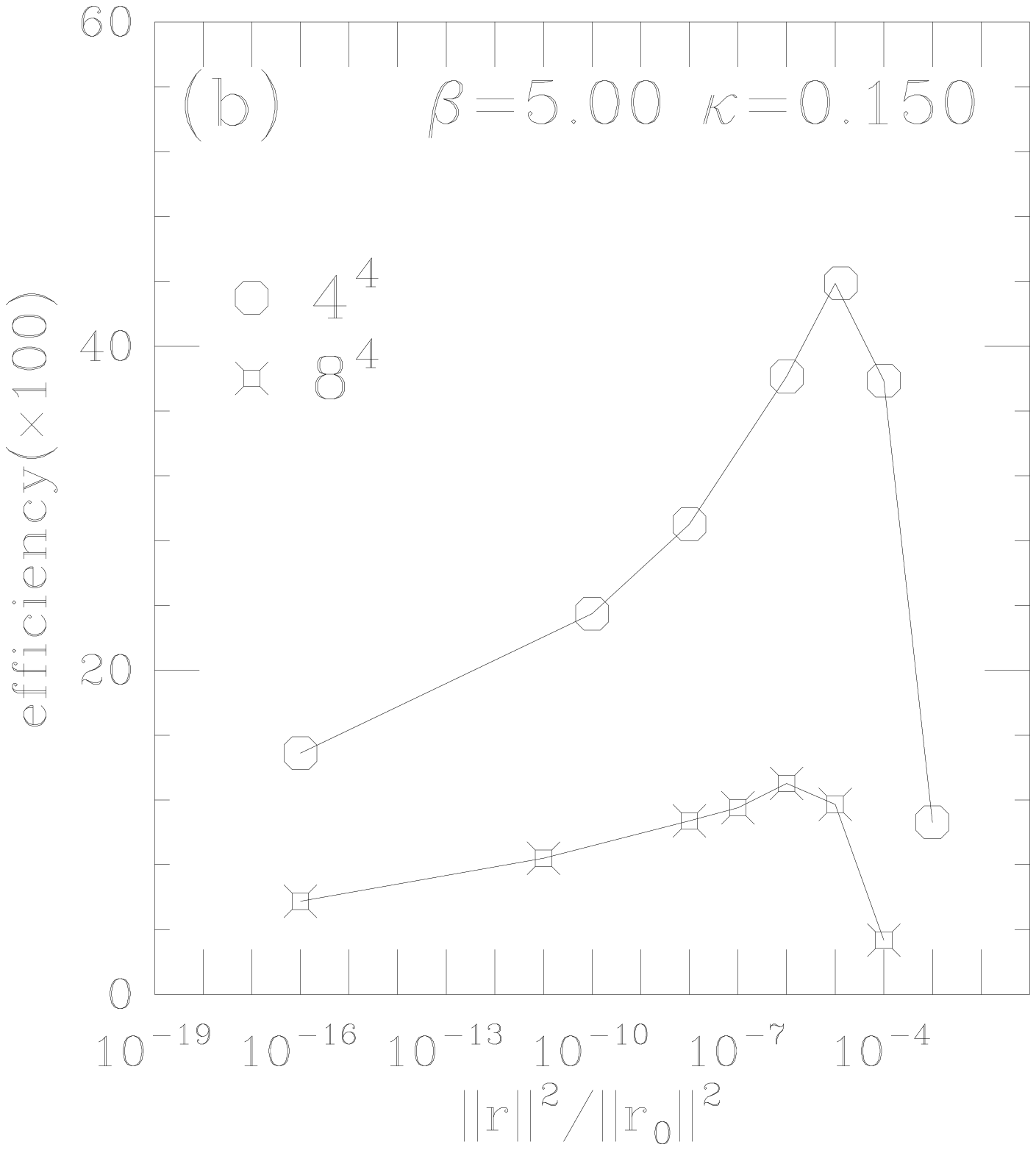,height=3cm,width=4.cm}}}
\caption{(a) Efficiency vs. step-size $\Delta t$ for polynomials of various 
degree $n$. For the standard HMC, the number of iterations in 
the $BiCG\gamma_5$ solver was taken as $n$. The stopping accuracy was set to 
$||r||=10^{-8}$,
where $||r||$ is the residual norm.
(b) Efficiency vs. residual norm. The trajectory length is  1(0.5) 
for a $4^4(8^4)$ lattice.}
\label{fig:chebyshev}
\end{figure}

L\"uscher proposed to approximate the inverse of the fermion matrix $D$ by
a Chebyshev polynomial $P_n(D)$\cite{Luscher} 
of degree $n$ with zeroes ${z_k}$: 
\begin{equation}
D^{-1} \simeq P_{n} (D) = c_n \Pi_{k=1}^n (D-z_k)
\end{equation}\label{Cheb}
This approximation allows the construction of a local update algorithm with
$n$ auxiliary bosonic fields\cite{Luscher}.  
Here we use the same approximation for the inverse of the fermion matrix 
which appears in the molecular dynamics Hamiltonian: 
\begin{equation}
\phi^\dagger (D D^\dagger)^{-1} \phi \rightarrow \phi^\dagger P_{n} (D)^\dagger  P_{n} (D)\phi
\end{equation}
This formulation does not require any linear solver. 
The fermionic force evaluation requires instead a predetermined
$N_D = 2 n$ multiplications by $D$. As $n$ is reduced however,
the approximation (\ref{Cheb}) increases the energy violation, 
which reduces the Metropolis acceptance. 
Fig.\ref{fig:chebyshev}(a) shows the efficiency $2/C_t$ vs. $\Delta t$. 
We find that the optimal step-size depends little on the degree $n$
of the polynomial. For the optimal $n$, the work is reduced by about 4
over standard HMC.

\subsection{Low accuracy stopping criterion}
Usually, in the molecular dynamics steps, the linear system which gives 
the fermionic force is solved to high accuracy.
The initial guess to the solution of this linear system
can then be extrapolated from previous results without measurably spoiling
time-reversibility. In this way, the number of iterations in the solver 
can be reduced by a factor $\sim$ 2\cite{solver}.  
On the other hand, it is also possible to reduce the accuracy in the solver,
provided that the solution is time-reversible. 
This is achieved simply by forfeiting the benefits of time extrapolation,
and taking always the same initial guess for the solver. While this
second approach is well-known, its benefits have not been studied 
to a comparable extent.

Here we systematically investigate the optimal accuracy, which minimizes 
the cost of HMC, by varying the stopping criterion in the solver.  
In this case, the actual form of the corresponding Hamiltonian $H_{MD}$
can not be known. 
Fig.\ref{fig:chebyshev}(b) shows efficiency, acceptance and number of iterations.
The number of iterations linearly decreases with the logarithm of the stopping 
criterion.
On the other hand, the acceptance remains practically constant up to a certain stopping criterion and 
then rapidly decreases. We measure the efficiency by $Acc/\#\; of\; iterations$.
The figure clearly shows that a low accuracy stopping criterion ($||r||^2/||r_0||^2\approx 10^{-5}$ and 
$10^{-6}$ on $4^4$ and $8^4$ lattices, respectively) is most efficient. 
However the gain over standard HMC ($||r||/||r_0||=10^{-8}$) appears to decrease as the volume increases.

\section{HMC with preconditioned action}

\begin{figure}[htb]
\centerline{\psfig{figure=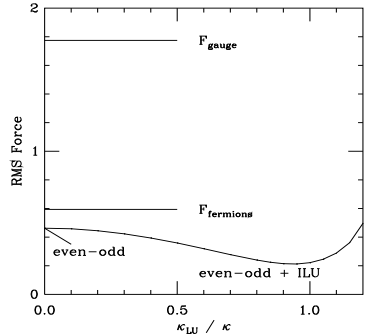,height=2.5cm,width=5cm}}
\caption{RMS force for various actions. }
\label{fig:force}
\end{figure}

The use of preconditioners is well-known for the solution of linear systems,
like the one which gives the fermionic force $F_f$ at each step. On the other
hand, little attention has been paid to preconditioning the fermionic 
action itself. One exception is even-odd preconditioning:
\begin{equation} 
S_f=\phi^\dagger_e(D_{ee}^\dagger D_{ee})^{-1}\phi_e 
\label{sf}
\end{equation}
where 
\begin{equation}
\det(1-\kappa M) = \det(1-\kappa^2 M_{eo}M_{oe})\equiv\det(D_{ee}).
\end{equation}
This formulation is widely used to reduce memory requirements and work per
force evaluation. What has not been noticed is that the fermionic force
from (\ref{sf}) is smaller than in the non-preconditioned case 
(see Fig.\ref{fig:force}),
allowing for an increase in the step-size without loss of acceptance.
This benefit stems from the better conditioning of $D_{ee}$ 
(to order $\kappa^2$) than $D$ (to order $\kappa$).
This preconditioning can be pushed to higher order with virtually no overhead.
Consider the partitioning of $D_{ee}$:
\begin{equation}
D_{ee}=1-(\kappa^2 M_{eo}M_{oe})_{+} - (\kappa^2 M_{eo}M_{oe})_{-},
\end{equation}
where +(-) stands for the lower(upper) triangular part. The lower(upper)
triangular matrices $L$ and $U = 1 - (\kappa^2 M_{eo}M_{oe})_{\pm}$ 
have determinant $1$. The preconditioned matrix 
$L^{-1}D_{ee}U^{-1} = 1+{\cal O}(\kappa^4)$ can then be substituted to $D_{ee}$ in 
the fermionic action:
\begin{eqnarray} 
S_f & = & \phi_e^\dagger[(L^{-1}D_{ee}U^{-1})^\dagger(L^{-1}D_{ee}U^{-1})]^{-1} \phi_e \nonumber \\
    & = & \phi_e^\dagger(UD_{ee}^{-1}L)(L^\dagger D_{ee}^{\dagger -1}U^\dagger) \phi_e.
\label{sfLU}
\end{eqnarray}
The corresponding fermionic force is obtained as:
\begin{equation}
\frac{\partial S_f}{\partial A_{\mu}} = 
\eta_e^{\dagger} \frac{\partial D_{ee}}{\partial A_{\mu}} x_e +
\phi_e^{\dagger} \frac{\partial U}{\partial A_{\mu}} x_e +
\eta_e^{\dagger} \frac{\partial L}{\partial A_{\mu}} L^{\dagger} \eta_e + h.c.
\end{equation}
with $\eta_e = D_{ee}^{\dagger -1} U^{\dagger} \phi_e$; 
$x_e = D_{ee}^{-1} L L^{\dagger} \eta_e$.
Since no inverse of $L$ or $U$ appears in this formulation, the overhead in the
force calculation is negligible. 
As shown in Fig.\ref{fig:force}, the fermionic force is reduced by a factor
$\sim 2$ as $\kappa_{LU}$ changes from $0$ (standard even-odd) to its
optimum slightly below $\kappa$. It becomes much smaller than the gauge force.
A Sexton-Weingarten integration scheme \cite{SW} then allows to take
very large fermionic step-sizes.

This $ILU$ preconditioner can be implemented to higher order in $\kappa$, with
some programming headache. The same idea can also be used for staggered 
fermions.

We tested $ILU$ preconditioning eq.(\ref{sfLU}) on a $16^3 \times 32$ lattice, at 
parameters studied by the SESAM collaboration ($\beta=5.6, \kappa=0.1575$)
\cite{SESAM}. We could increase the fermionic step-size from $0.01$
(SESAM) to $0.05$ while maintaining a $66\%$ acceptance. A more extended,
though still preliminary test was conducted on an $8^4$ lattice at $\beta=5.6,
\kappa=0.1585$. These parameters were chosen in \cite{Jansen} to compare
HMC with the multiboson method, very near criticality. $ILU$ preconditioning
allowed us to increase the fermionic step-size from $0.075$ to $0.12$. At
the same time, we used a low-accuracy stopping criterion $||r||/||r_0||=10^{-3}$
during the trajectory, reducing the number of matrix multiplications per step
from $\sim 2\times159$ to $\sim 2\times66$. The autocorrelation time
of the plaquette seems unaffected by preconditioning.

If these preliminary indications are confirmed, the prospect is to improve the
efficiency of HMC by a factor ${\cal O}(5)$.

\end{document}